\documentclass[prl,twocolumn,showpacs,superscriptaddress,floatfix,amssymb,nofootnoteinbib]{revtex4}

\usepackage[pdftex]{graphicx}
\usepackage{dcolumn}
\usepackage[colorlinks]{hyperref} 
\usepackage{bm}
\usepackage{amsfonts,amssymb,amsmath}        
\graphicspath{{./figs/}}

\newcommand{\tr}{\mbox{tr}}
\newcommand{\ket}[1]{\left | #1 \right \rangle}
\newcommand{\bra}[1]{\left \langle #1 \right |}

\newcommand{\beq}{\begin{equation}}
\newcommand{\eeq}{\end{equation}}
\newcommand{\beqa}{\begin{eqnarray}}
\newcommand{\eeqa}{\end{eqnarray}}

\newcommand{\avg}[1]{\left\langle\langle #1 \right\rangle\rangle}

\begin{document}

\title{Transport properties of anyons in random topological environments}
\author{V. Zatloukal}
\affiliation{Faculty of Nuclear Sciences and Physical Engineering, Czech Technical University in Prague, B\v{r}ehov\'{a} 7, 115 19 Praha 1, Czech Republic}
\author{L. Lehman}
\affiliation{Centre for Engineered Quantum Systems, Department of Physics and Astronomy, Macquarie University, North Ryde, NSW 2109, Australia}
\author{S. Singh}
\affiliation{Centre for Engineered Quantum Systems, Department of Physics and Astronomy, Macquarie University, North Ryde, NSW 2109, Australia}
\author{J.K. Pachos}
\affiliation{School of Physics and Astronomy, University of Leeds, Leeds LS2 9JT, UK}
\author{G.K. Brennen}
\affiliation{Centre for Engineered Quantum Systems, Department of Physics and Astronomy, Macquarie University, North Ryde, NSW 2109, Australia}
\affiliation{School of Physics and Astronomy, University of Leeds, Leeds LS2 9JT, UK}


\date{\today}  


\begin{abstract}

The quasi one-dimensional transport of Abelian and non-Abelian anyons is studied in the presence of a random topological background. In particular, we consider the quantum walk of an anyon that braids around islands of randomly filled static anyons of the same type. Two distinct behaviours are identified. We analytically demonstrate that all types of Abelian anyons localise purely due to the statistical phases induced by their random anyonic environment. In contrast, we numerically show that non-Abelian Ising anyons do not localise. This is due to their entanglement with the anyonic environment that effectively induces dephasing. Our study demonstrates that localisation properties strongly depend on non-local topological interactions and it provides a clear distinction in the transport properties of Abelian and non-Abelian statistics.

\end{abstract}

\maketitle


In systems with physics constrained to two dimensions, point like particles named anyons can occur which have more general statistics than bosons or fermions \cite{Leinass}. Beyond mere possible existence they were found to be a good description for low lying quasi-particle excitations of fractional quantum Hall systems, Majorana edge modes of nanowires, and they exactly describe excitations in various strongly correlated two dimensional spin lattice models \cite{Pachos}.  Recently there has been experimental progress in preparation and control of systems capable of exhibiting topological order with the goal to observe anyonic statistics \cite{willett,Mourik}. This is further motivated by the discovery that braiding some types of non-Abelian anyons provides for naturally fault tolerant quantum computing \cite{Freedman,Pachos}.  
As we are not yet able to manipulate anyons individually, it would be beneficial to reveal their exotic  braiding properties on macroscopic scale. In particular, we are interested in the transport properties of anyons and their possible localisation, that can have direct observable consequences. 

Transport properties of anyons in ordered backgrounds were studied in  \cite{Brennen}. There a discrete-time quantum walk was considered of one mobile anyon braiding around a canonically ordered set of static anyons positioned on a line. While for Abelian anyons the dispersion of the walker is quadratic, as in usual quantum walk \cite{Ambainis}, the non-Abelian anyons have richer behaviour induced by their braiding properties. Specifically, it was shown in \cite{Lehman} that the Ising non-Abelian anyons exhibit asymptotically a linear dispersion relation, characteristic to classical random walks. Entanglement resulting from braiding the mobile non-Abelian anyon around the static ones is sufficient to suppress quantum correlations responsible for the quadratic speed-up.

In this letter, we investigate the role of disorder on the propagation of both, Abelian and non-Abelian, anyons. It has been known for more than five decades that randomised local potentials can suppress diffusion of quantum particles --- a phenomenon known as Anderson localisation \cite{Anderson}. This mechanism is based on randomisation of phases that correspond to individual particle histories and consequent destructive interference. Here we consider a discrete-time anyonic quantum walk in disordered topological background. In particular, we consider a walker that braids around islands canonically arranged on a line, where the number of static anyons at a given island assumes a random value. For the Abelian anyons this causes the walker acquiring random {\em discrete} phases which, as we demonstrate, leads to localisation. For the non-Abelian ones we resort to numerical treatment. A continuous-time analog, studied numerically with the matrix product state approach, concludes that Ising non-Abelian anyons do not localise. These generic behaviours provide a very clear distinction between Abelian and non-Abelian anyons. 

\begin{figure}
\begin{center}
\includegraphics[width=1\columnwidth]{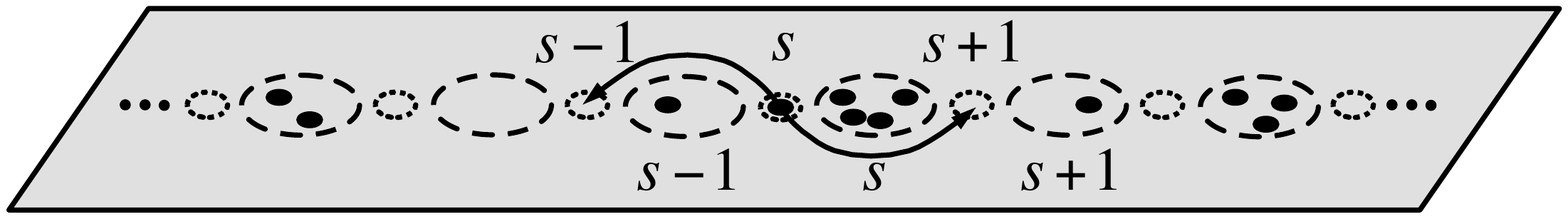}
\end{center}
\caption{The quasi one-dimensional quantum walk of an anyon braiding counterclockwise around islands filled with a random number of static anyons of the same type. The islands, denoted with dashed circles, are canonically arranged on the line. The possible positions of the walker are denoted by dotted circles placed in between the islands.}
\label{fig:Figure1}
\end{figure}

{\em The discrete-time model:-}
Our setup consists of $n$ ``islands" canonically ordered on the surface and labeled by index $s$, as shown in Fig. \ref{fig:Figure1}. The $s$-th island is occupied by $m_s$ static anyons $(m_s \geq 0)$ and the configuration, represented by vector $\vec{m} = (m_1,\dotsc,m_n)$, is supposed to be fixed during the course of the walk. Anyons are labeled within an island from left to right by an index $i_s = 1,\dotsc,m_s$. The mobile \emph{walker} anyon hops between neighbouring sites winding counterclockwise around the islands. Hence our system is quasi one-dimensional. We denote possible walker's spatial positions also by $s$, with the convention that position $s$ lies between islands $s-1$ and $s$, as shown in Fig. \ref{fig:Figure1}. The distance between sites is set to unity, though for purely topologically interactions the distance scale is irrelevant. Hopping direction is controlled by the coin state: $\ket{0}$ moves the walker to the left, $\ket{1}$ to the right. The total Hilbert space decomposes as $\mathcal{H} = \mathcal{H}_{\rm space} \otimes \mathcal{H}_{\rm coin} \otimes \mathcal{H}_{\rm fusion}$, where $\mathcal{H}_{\rm space} = {\rm span}\{ \ket{s} \}_{s=1}^n$, $\mathcal{H}_{\rm coin} = {\rm span}\{ \ket{0},\ket{1} \}$, and $ \mathcal{H}_{\rm fusion}$ enumerates all distinct measurement outcomes of topological charge when pairs of anyons are fused together \cite{Freedman}.

One step of the walk is defined as a composition of two unitary operations, $W = T U$, where $U = \frac{1}{\sqrt{2}}\left( \begin{smallmatrix} 1&1\\1&-1 \end{smallmatrix} \right)$ acts in the coin space, and $T$ is a conditional braiding operator which moves the walker left or right depending on the coin state:
\begin{equation}
T = \sum_{s=1}^{n} \ket{s-1}\bra{s} \otimes \ket{0}\bra{0} \otimes \hat{b}_{s-1} + \ket{s+1}\bra{s} \otimes \ket{1}\bra{1} \otimes \check{b}_{s} ~,
\end{equation}
where
$\hat{b}_{s} = b_{s,1} \dotsm b_{s,m_s}$, $\check{b}_{s} = b_{s,m_s} \dotsm b_{s,1}$, and $\hat{b}_{s} = \check{b}_{s} = 1$ if $m_s=0$. The operators $\{b_{s,i_s}\}$, acting on the fusion space $\mathcal{H}_{\rm fusion}$, form a unitary representation of the $r$-strand braid group, $r=1+\sum_{s=1}^{n} m_s$, which reflects the type of anyons we choose. To make $T$ unitary, we assume periodic boundary conditions ($\ket{0}_{\rm space} = \ket{n}_{\rm space}$) but will be concerned with walks satisfying $n/2 < t$, where $t$ is the number of steps, so that winding around the surface is not an issue.

Let the system's initial state be $\ket{\Psi(0)} = \ket{s_0} \ket{c_0} \ket{\Phi_0}$, where $s_0 = \lceil n/2 \rceil$ is the initial position of the walker, $c_0 = 0$ denotes the initial state of the coin, and $\Phi_0$ depends on the initial state of the anyons. After $t$ iterations of the one step operator $W$, the state becomes $\ket{\Psi(t)} = W^t \ket{\Psi(0)}$ --- a superposition over all coin histories $\vec{a} \in \{0,1\}^{\otimes t}$, weighted by appropriate phase factors. The reduced state of the spatial degree of freedom of the walker is
\begin{align}
\rho_{\rm space}(t) &= {\rm tr}_{\rm coin} {\rm tr}_{\rm fusion} \ket{\Psi(t)}\bra{\Psi(t)} \nonumber\\
&= \sum_{\vec{a},\vec{a}'} 
\tr \mathcal{U}_{\vec{a}\vec{a}'} 
\tr \mathcal{Y}_{\vec{a}\vec{a}'}
\ket{s_{\vec{a}}} \bra{s_{\vec{a}'}} ~,
\end{align}
where $s_{\vec{a}} = s_0 + \sum_{k=1}^{t} (2 a_k - 1)$ is the walker's final position corresponding to the coin history $\vec{a} = (a_1,\dotsc,a_t)$; 
$\tr \mathcal{U}_{\vec{a}\vec{a}'} = \frac{1}{2^t} (-1)^{z(\vec{a},\vec{a}')}$, with $z(\vec{a},\vec{a}') \equiv \sum_{k=1}^{t-1} (a_k a_{k+1} + a'_k a'_{k+1})$, is a partial trace over the coin DOF; and $\mathcal{Y}_{\vec{a}\vec{a}'} = B_{\vec{a}} \ket{\Phi_0}\bra{\Phi_0} B_{\vec{a}'}^\dag$ acts in the fusion space. The braid word $B_{\vec{a}}$ can be constructed recursively from a given coin history $\vec{a}$:
\begin{equation}
B_{\vec{a}^{(k+1)}} = \left\{\begin{array}{ll}
\hat{b}_{s_{\vec{a}^{(k)}}-1} B_{\vec{a}^{(k)}} & \textrm{if}~~ a_{k+1} = 0 \\ \check{b}_{s_{\vec{a}^{(k)}}} B_{\vec{a}^{(k)}} & \textrm{if}~~ a_{k+1} = 1
\end{array}\right. ~,
\end{equation}
where $\vec{a}^{(k)} = (a_1,\dotsc,a_k)$ is a truncation of $\vec{a}$, $s_{\vec{a}^{(0)}} = s_0$, and $B_{\vec{a}^{(0)}} = 1$.

The spatial distribution of the walker after $t$ steps is given by diagonal elements of the reduced density matrix, 
\begin{equation} \label{fixed configuration distribution}
p_{\vec{m}}(s,t) \equiv \bra{s} \rho_{\rm space}(t) \ket{s} 
= \frac{1}{2^t} \sum_{(\vec{a},\vec{a}') \rightsquigarrow s} (-1)^{z(\vec{a},\vec{a}')} \tr \mathcal{Y}_{\vec{a}\vec{a}'} ~,
\end{equation}
where ``$(\vec{a},\vec{a}') \rightsquigarrow s$" denotes the set of pairs of paths $(\vec{a},\vec{a}')$ satisfying $a_t=a'_t$ and $s_{\vec{a}} = s_{\vec{a}'} = s$. The subscript $\vec{m}$ indicates a fixed island occupation configuration. Variance of the probability distribution $p_{\vec{m}}(s,t)$ is defined in the usual way: $\sigma^2_{\vec{m}}(t) = \sum_s p_{\vec{m}}(s,t) s^2 - \left( \sum_s p_{\vec{m}}(s,t) s \right)^2$.

In the studies of transport phenomena in disordered environments, one is interested in quantities that result from averaging over all random background configurations. We shall assume that the island occupation numbers $m_s$ are independent and identically distributed random variables with distribution $W(m_s)$. The probability of occurrence of a configuration $\vec{m}$ is then simply $W_{\vec{m}} = \prod_{s=1}^n W(m_s)$, and we denote configuration average of a quantity $Q_{\vec{m}}$ by $\langle\langle Q \rangle\rangle \equiv \sum_{\vec{m}} W_{\vec{m}} Q_{\vec{m}}$. The average position distribution after a $t$-step walk is given by
\begin{equation} \label{average position distribution}
\langle\langle p(s,t)\rangle\rangle = \frac{1}{2^t} \sum_{(\vec{a},\vec{a}') \rightsquigarrow s} (-1)^{z(\vec{a},\vec{a}')} \langle \langle \tr\mathcal{Y}_{\vec{a}\vec{a}'}\rangle\rangle ~.
\end{equation}
The topological quantity $\langle\langle \tr \mathcal{Y}_{\vec{a}\vec{a}'}\rangle\rangle$, that depends on particle statistics, governs the transport of an anyonic walker in random background.

{\em  Abelian anyons:-}
For Abelian anyons the braid generators $\{b_{s,i_s}\}$ are all equal to $e^{i \phi}$. We assume the anyonic exchange angle to be $\phi = \pm\frac{\pi}{N}$, $N \in \mathbb{N}$. The fusion space is one-dimensional, $\mathcal{H}_{\rm fusion} \simeq \mathbb{C}$, and we can choose $\ket{\Phi_0}$ arbitrarily. Upon introducing the linking numbers
$
\ell_s(\vec{a},\vec{a}') = \frac{\#(\hat{b}_s {\rm ~and~} \check{b}_s {\rm ~in~} B_{\vec{a}}) - \#(\hat{b}_s^\dag {\rm ~and~} \check{b}_s^\dag {\rm ~in~} B_{\vec{a}'}^\dag)}{2} ~,
$
that count the number of times the walker's trajectory $(\vec{a},\vec{a}')$ winds around an island $s$, $\tr \mathcal{Y}_{\vec{a}\vec{a}'}$ reduces to $\prod_{s=1}^n e^{\pm i {2 \frac{\pi}{N}} m_s \ell_s(\vec{a},\vec{a}')}$.  

\begin{figure}
\includegraphics[width=\columnwidth]{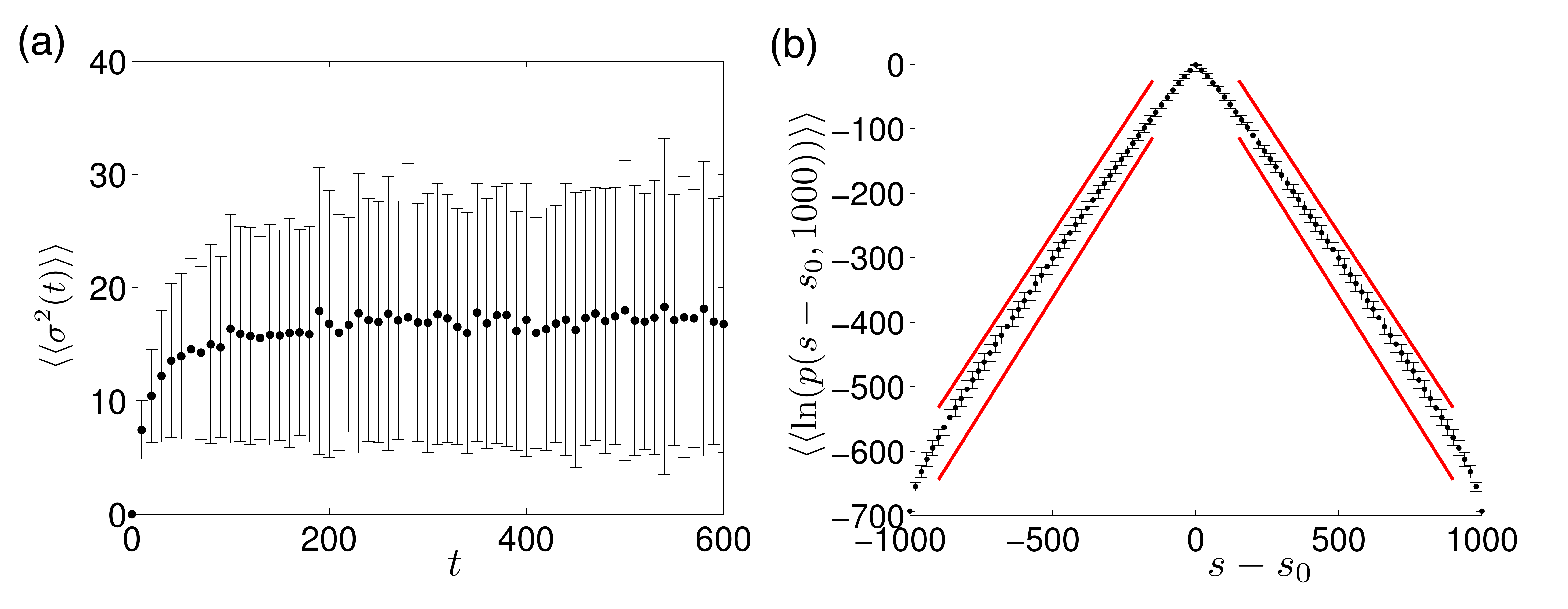}
\caption{Numerical results for localisation of Abelian anyons. The exchange statistics is $\phi=\frac{\pi}{8}$ and the statistics is averaged over a random background of island occupations where the distribution in each is uniform over $m_s\in\{0,\ldots,7\}$.
(a) Average variance as a function of time $t$ for up to 600 time steps.
The averages are taken over at least 500 charge configurations. (For clarity, only every 10th step is plotted.)
(b) Average of the logarithm of probability distribution at time step $t=1000$, taken over 10000 charge configurations.
Red lines correspond to bounds for the localisation length given in Eq. (\ref{localization length estimates}).
(Only the relevant region $-700\leq s-s_0\leq700$ and every 20th site are plotted.)
In both a) and b), the error bars are given by standard deviation.}
\label{fig:abelian}
\end{figure}

For simplicity we consider uniform occupation distribution: $W(m) = 1/N$ for $1 \leq m \leq N$. Then the average position distribution of the walker after $t$ steps, $\langle\langle p^{(\pm\frac{\pi}{N})}(s,t)\rangle\rangle$, is given by (\ref{average position distribution}) with
\begin{equation} \label{average trace Abelian}
\langle\langle\tr\mathcal{Y}_{\vec{a}\vec{a}'}^{(\pm\frac{\pi}{N})}\rangle\rangle
= \prod_{s=1}^n \delta_{0,\ell_s(\vec{a},\vec{a}') ~{\rm mod}~ N} ~,
\end{equation}
where $\delta_{i,j}$ is the Kronecker symbol. Fig. \ref{fig:abelian} shows numerical results of $\avg{p^{(\pm\frac{\pi}{8})}(s,t)}$ for large time $t$ and time behavior of the average variance $\avg{\sigma^2{^{(\pm\frac{\pi}{8})}(t)}}$. Furthermore, we find that the results are essentially insensitive to the choice of $N$ (except for the case $N=1$ which corresponds to fermions, and which thus reduces to a usual quantum walk). The variance approaches a constant value and the asymptotic position distribution assumes a characteristic exponential shape, $\avg{p^{(\pm\frac{\pi}{N})}(s,t \rightarrow \infty)} \sim e^{-\frac{|s-s_0|}{\xi_{\rm loc}}}$, with the localization length $\xi_{\rm loc} \doteq 1.44$. This is a clear manifestation of the Anderson localization of Abelian anyons.

Theoretical explanation of the above numerical results is based on the correspondence between our Abelian anyonic quantum walk in disordered background and the one-dimensional multiple scattering model presented in \cite{Muller}. In the latter, scatterers are arranged in line with random distances between neighbours (see Appendix). Incoming light undergoes a series of scattering events and eventually localises due to the randomness in phases that individual trajectories accumulate during their passage between consecutive scatterers. The spatial randomness can be translated into a disorder in the population of the islands in the anyonic quantum walk model. Furthermore, reflection and transmission coefficients of the scatterers can be identified with the entries of the coin operator. The scatterer distances take discrete and finite many  values reflecting the discrete nature in the populations of the islands. Localisation of the walking Abelian anyon then follows, with the localisation length $\xi_{\rm loc} \approx 1/\ln 2 \doteq 1.443$. More precisely, in Appendix \ref{sec:scattering model} we derive the bound
\begin{equation} \label{localization length estimates}
\frac{1}{\ln 2 + \ln (1+2^{N/2})^{2/N}} \leq \xi_{loc} \leq \frac{1}{\ln 2 + \ln (1-2^{N/2})^{2/N}} ~,
\end{equation}
where the anyonic statistical angle $\phi = \frac{\pi}{N}$. This theoretical result is in agreement with exact numerical treatment presented in Fig. \ref{fig:abelian}.

{\em Ising non-Abelian anyons:- }
We consider the Ising model of non-Abelian anyons. There are three kinds of particles in this model: vacuum, fermion, and non-Abelian anyon. In analogy with \cite{Lehman}, the initial fusion state $\ket{\Phi_0}$ describes the vacuum configuration of pairs of anyons with half the members braided to the right. Here, they randomly populate islands (except for the walker) to form a disordered background configuration $\vec{m}$. The braid generators $\{b_{s,i_s}\}$ are now unitary matrices, the dimension of which grows like $d^{|\vec{m}|}$, where $|\vec{m}| \equiv \sum_{s=1}^n m_s$, and $d = \sqrt{2}$ is the \emph{quantum dimension} of Ising anyons. 

The trace over the fusion degree of freedom can be related to the Kauffman bracket polynomial $\langle L_{\vec{a}\vec{a}'}\rangle$ \cite{Kauffman} of a link $L_{\vec{a}\vec{a}'}$, which arises from the Markov closure of the braid word $B_{\vec{a}'}^\dag B_{\vec{a}}$ \cite{Lehman, Brennen, Aharonov}. Moreover, $\langle L_{\vec{a}\vec{a}'}\rangle$ can be expressed in terms of the Jones polynomial $V_{L_{\vec{a}\vec{a}'}}(q)$ \cite{Jones}. Altogether,
\begin{equation}
\label{JonesDef}
\tr \mathcal{Y}_{\vec{a}\vec{a}'} 
= \frac{\langle L_{\vec{a}\vec{a}'}\rangle(q^{-1/4})}{d^{|\vec{m}|}}
= \frac{\left(- q^{-3/4}\right)^{w(L_{\vec{a}\vec{a}'})} V_{L_{\vec{a}\vec{a}'}}(q)}{d^{|\vec{m}|}} ~,
\end{equation}
where the \emph{writhe} $w(L_{\vec{a}\vec{a}'}) =  2 \sum_{s=1}^n m_s \ell_s(\vec{a},\vec{a}')$. For Ising anyons, which correspond to spin-$1/2$ irreps of the quantum group $SU(2)_2$ we have specifically $q=i$. In this case $V_{L_{\vec{a}\vec{a}'}}(i)$ can be further simplified in terms of simple topological characteristics of the link $L_{\vec{a}\vec{a}'}$ (see Appendix). Our final result for $\tr \mathcal{Y}_{\vec{a}\vec{a}'}$ reads
\begin{equation} \label{Ising fusion trace}
\tr \mathcal{Y}_{\vec{a}\vec{a}'} = \prod_{\substack{s=1 \\ m_s>0}}^n \widetilde{\ell}_s (-i)^{\frac{\ell_s}{2} m_s} \prod_{1 \leq s' < s'' \leq n} (-1)^{m_{s'} m_{s''} \tau(s',s'')} ~,
\end{equation}
where $\widetilde{\ell}_s \equiv \delta_{0,\ell_s ~{\rm mod}~ 2}$ and $\tau(s',s'')$ is the Milnor triple invariant of a three-component sublink of $L_{\vec{a}\vec{a}'}$ formed by strands corresponding to the walker and the islands $s'$ and $s''$. (We do not explicitly indicate the $(\vec{a},\vec{a}')$-dependence of the right hand side.) For the case of uniform background configuration we recover the case studied in  \cite{Lehman}.

To illustrate the role of disorder in a non-Abelian anyonic quantum walk, we choose uniform island occupation probabilities. As (\ref{Ising fusion trace}) is for $m_s>0$ $4$-periodic in $m_s$ we take $W(m) = 1/4$ for $1 \leq m \leq 4$ and $W(m) = 0$ otherwise \cite{footnote}. The configuration average of (\ref{Ising fusion trace}) can be partially carried through,
as shown in Appendix. The average position distribution of the Ising quantum walk after $t$ steps, $\avg{p^{\rm (Ising)}(s,t)}$, is given by (\ref{average position distribution}) with average fusion space trace
\begin{equation} \label{average trace Ising}
\langle\langle \tr\mathcal{Y}_{\vec{a}\vec{a}'}^{\rm (Ising)} \rangle \rangle=  \mathcal{T}_{\vec{a}\vec{a}'} \prod_{s=1}^n \delta_{0,\ell_s(\vec{a},\vec{a}') ~{\rm mod}~ 8} ~,
\end{equation}
where 
$\mathcal{T}_{\vec{a}\vec{a}'} = \frac{1}{2^n} \sum_{\vec{m} \in \{0,1\}^n} \prod_{1\leq s'<s'' \leq n} (-1)^{m_r m_s \tau(s',s'')}$ can be interpreted as an arithmetic mean of the quantity $(-1)^{arf(L^{*}_{\vec{a}\vec{a}'})}$ taken over all sublinks $L^{*}_{\vec{a}\vec{a}'}$ of a link $L_{\vec{a}\vec{a}'}$.
On comparing (\ref{average trace Ising}) to the Abelian expression (\ref{average trace Abelian}) for $N=8$,
they are identical except for the prefactor $\mathcal{T}_{\vec{a}\vec{a}'}$. Considering
Eq. (\ref{average trace Abelian}) as the coherent expression where the quantum interference of probability amplitudes
causes localisation, the $\mathcal{T}_{\vec{a}\vec{a}'}$ coefficient can be viewed as a noise term which
might or might not destroy the interference. In the Supplementary material, we argue that at short time scales,
this term does not preserve memory and introduces temporal randomness. By results of Ref. \cite{spatiotemporal},
localisation does not occur in the presence of both spatial and temporal randomness, therefore we conjecture
that non-Abelian anyons do not localise in the asymptotic limit $t\rightarrow\infty$.

\begin{figure}
\begin{center}
\includegraphics[width=\columnwidth]{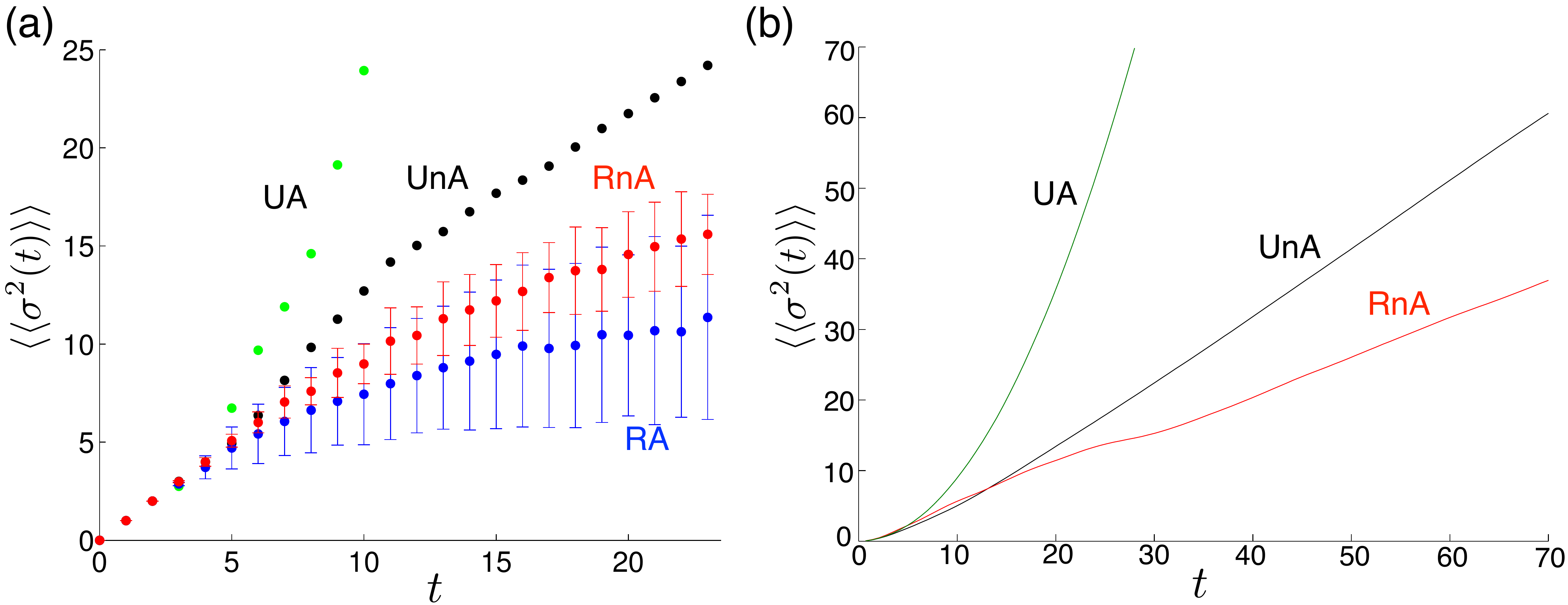}
\end{center}
\caption{Numerical results for transport of Abelian and non-Abelian anyons.  (a) 
Exact results for the discrete time quantum walk with anyons over an $n=46$ sized lattice.  Here is shown:
variance of an Abelian anyon  with $\frac{\pi}{8}$ exchange statistics around a uniform background of islands singly occupied by Abelian anyons (Uniform Abelian=UA), variance of a Ising model anyon around a uniform background $m_s=1 \forall s$ of Ising anyons (Uniform non-Abelian=UnA), variance of an Ising model anyon around a random background of islands where the distribution in each is uniform over $m_s\in\{0,\ldots,4\}$ Ising anyons and averaging is taken over 50 background configurations (Random non-Abelian=RnA), and variance of the Abelian anyon averaged over at least $100$ background configurations of Abelian anyons with occupation probability uniform over $m_s\in\{0,\ldots,7\}$ (Random Abelian=RA). (b)
Variance of anyons in a Hubbard model on a ladder realising a continuous time anyonic quantum walk over an $n=100$ sized lattice.  Here space and time axes are scaled so that a continuous time classical diffusion would have diffusion coefficient providing $\sigma^2(t)=1$. 
The asymptotic slope for the case RnA averaged in the same way as in (a) is $0.5426$ and the slope remains positive and less than one within one sigma variance (error bars suppressed for clarity).  Numerics are obtained using an approximate method employing real time evolution of an anyonic MPS with bond dimension equal to 100.}
\label{fig:variancenonab1}
\end{figure}

We have also performed numerical calculations for non-Abelian anyons using two methods.
First, the probability distribution was calculated using Eqs. (\ref{fixed configuration distribution}) and
(\ref{Ising fusion trace}) up to 23 time steps (see inset of Fig. \ref{fig:variancenonab1}), but it is hard to
determine from these results whether localisation occurs or not. To obtain results for longer time scales,
we used a Hubbard model \cite{sukhi} on a ladder realising a continuous time anyonic quantum walk for an $n=100$ sized lattice with open (reflecting) boundaries. These results were obtained by using the ``Time-Evolving Block Decimation'' (TEBD) \cite{Vidal04} algorithm that is based on Matrix Product States (MPS). Fig. \ref{fig:variancenonab1} shows that for non-Abelian anyons the variance grows linearly as a function of time, indicating no signature of localisation. 


{\em Conclusions:-}
A new characteristic of statistical behaviour has been presented in terms of the localisation properties of anyons. We demonstrated that Abelian anyons propagating in an environment of other anyons of the same type in random positions, localise, regardless of their mutual anyonic statistics. To prove this we demonstrated that localisation can occur not only in the presence of a continuous random variable, but also of discrete, where the localisation length is tightly determined. For the non-Abelian case we considered the Ising anyonic model. We showed that it fails to localise due to the entanglement of the walker with its environment that can be effectively described as dephasing of the walk leading to linear dispersion. This argument is supported by exact numerical results for a small number of walk steps and by an approximate numerical method that employs matrix product states. We expect delocalisation to occur even for more general $SU(2)_k$ non-Abelian anyons due to the entanglement of the walker with the environment. Indeed, in \cite{Lehman2} it has been demonstrated that the entanglement of the walker with its uniform environment for any $k>1$ is strong enough to cause the walk to decohere obtaining eventually a classical diffusive behaviour. 

Finally, our study provides a new paradigm of disorder that has the ability to override the localisation properties of quasiparticles with anyonic statistics. In a real system with random potentials Abelian anyons that do not localise due to the potentials can be localised just by the presence of a random anyonic environment. On the other hand, non-Abelian anyons that would localise due to some random potential can be delocalised if a random background of anyons of the same type is present.

We acknowledge helpful discussion with S. Simon and Z. Wang.  V.Z. received support from CTU grant 811320, G.K.B. received support from the ARC via the Centre of Excellence in Engineered Quantum Systems (EQuS), project number CE110001013 and from DP1094758, and J.K.P. received support from the EPSRC.


\appendix

\section{Multiple scattering model of an Abelian anyonic quantum walk}
 \label{sec:scattering model}
Propagation of Abelian anyons and waves that undergo multiple scattering on a series of randomly distant scatterers is governed by the same interference mechanism. We base correspondence between the two on physical intuition rather than mathematical rigor. 

\begin{figure}[h]
\begin{center}
\includegraphics[scale=0.7]{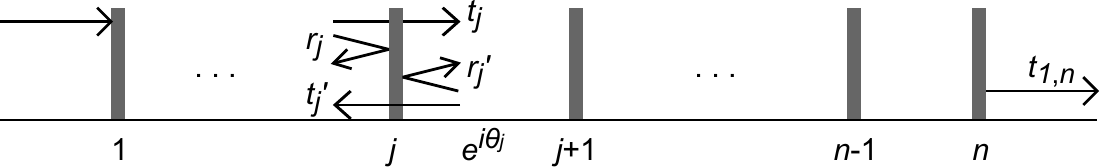}
\end{center}
\caption{In the multiple scattering model, the wave approaches (from the left) a series of $n$ scatterers, and is transmitted with the amplitude $t_{1,n}$. The scatterers are arranged in line with random distances between neighbours. Hence, the phases $e^{i \theta_j}$ that the wave acquires during travelling from a scatterer $j$ to $j+1$ are also random.  The complex quantities $r_j, t_j / r_j', t_j'$ are the reflection and transmission amplitudes for the wave impinging from the left / right.}
\label{fig:scatterers}
\end{figure}

The scattering model is described by Fig. \ref{fig:scatterers}. A monochromatic wave incident from the left scatters on a series of scatterers characterized by ``from left / from right" reflection and transmission coefficients $r_j, t_j / r_j', t_j'$. The distance between two successive scatterers $j$ and $j+1$ is random, such that the phase that the wave acquires when traveling between $j$ and $j+1$ is $e^{i\theta_j}$. Our result on localization of waves propagating through a random configuration of scatterers will be merely a discrete version of \cite{MullerII}, section $2$.

Denote by $t_{1,n}$ the block amplitude of transmission from the ``left of scatterer $1$" to the ``right of scatterer $n$"; and by $r_{1,n}'$ the reflection amplitude from the block ``$1$ to $n$" when approaching from the right. $t_{1,n}$ can be expressed by the series
\begin{eqnarray}
t_{1,n} &=& t_{1,n-1} e^{i \theta_{n-1}} \sum_{k=0}^\infty \left( r_{n} e^{i \theta_{n-1}} r_{1,n-1}' e^{i \theta_{n-1}} \right)^k t_{n} \nonumber\\
&=& \frac{t_{1,n-1} e^{i\theta_{n-1}} t_n}{1 - r_n r_{1,n-1}' e^{i 2 \theta_{n-1}}} ~.
\end{eqnarray}
The corresponding transmission probability and its logarithm are given by
\begin{equation}
|t_{1,n}|^2 = \frac{|t_{1,n-1}|^2 |t_n|^2}{| 1 - r_n r_{1,n-1}' e^{i 2 \theta_{n-1}} |^2} ~,
\end{equation}
\begin{equation} \label{log transmission probability recurrence}
\ln |t_{1,n}|^2 = \ln |t_{1,n-1}|^2 + \ln |t_n|^2 - \ln | 1 - r_n r_{1,n-1}' e^{i 2 \theta_{n-1}} |^2 ~.
\end{equation}
The reflection and transmission amplitudes $t_{1,n}$, $r_{1,n}'$ are random variables that depend on the configuration of scatterers $1, \ldots, n$, i.e. on the angles $\theta_1,\ldots,\theta_{n-1}$. We assume that $\theta_j$'s are identically distributed independent random variables with a uniform distribution over the discrete set $\{ \frac{\pi}{N} m \mid m=0,\ldots,N-1 \}$ ($\frac{\pi}{N}$ will be identified with the anyonic exchange angle $\varphi$). We shall denote by $\avg{\big(\ldots\big)}$ the statistical average over the angles $\theta_1,\ldots,\theta_{n-1}$, i.e.
\begin{equation}
\avg{\big(\ldots\big)} \equiv \sum_{m_1=0}^{N-1} \frac{1}{N} \ldots \sum_{m_{n-1}=0}^{N-1} \frac{1}{N} \big(\ldots\big) ~.
\end{equation}

Averaging of (\ref{log transmission probability recurrence}) leads to
\begin{align} \label{avg log transmission probability recurrence}
\avg{\ln |t_{1,n}|^2} 
&= \avg{\ln |t_{1,n-1}|^2} + \ln |t_n|^2 \nonumber\\
&~~~~~~ - \avg{\ln | 1 - r_n r_{1,n-1}' e^{i 2 \theta_{n-1}} |^2} ~.
\end{align}
To proceed, we carry out the $\theta_{n-1}$-average in the last term of (\ref{avg log transmission probability recurrence}),
\begin{align} \label{one theta averaging}
&\sum_{m_{n-1}=0}^{N-1} \frac{1}{N} \ln | 1 - r_n r_{1,n-1}' e^{i \frac{2 \pi}{N} m_{n-1}} |^2 \nonumber\\ 
&= \frac{1}{N} \ln \left| \prod_{m_{n-1}=0}^{N-1} ( 1 - r_n r_{1,n-1}' e^{i \frac{2 \pi}{N} m_{n-1}} ) \right|^2 \nonumber\\ 
&= \frac{1}{N} \ln | 1-(r_n r_{1,n-1}')^N |^2
\end{align}
In the latter equality we used the fact that
\begin{equation}
\prod_{m=1}^N ( 1 - C e^{i \frac{2 \pi}{N} m} ) = 1 - C^N ~,
\end{equation}
which can be proven by using the Newton's identities between elementary symmetric polynomials and power sums \cite{Macdonald}.

The $\theta_1,\ldots,\theta_{n-2}$-average of (\ref{one theta averaging}) becomes trivial once we estimate ($|r_{1,n-1}'| \leq 1$)
\begin{eqnarray}
\ln ( 1-(|r_n| |r_{1,n-1}'|)^N )^2 &\leq& \ln | 1-(r_n r_{1,n-1}')^N |^2\nonumber \\
&\leq& \ln ( 1+(|r_n| |r_{1,n-1}'|)^N )^2,\\
\ln ( 1-|r_n|^N )^2 &\leq& \ln | 1-(r_n r_{1,n-1}')^N |^2 \nonumber\\
& \leq& \ln ( 1+|r_n|^N )^2 ~.
\end{eqnarray}
The upper and lower bounds of relation (\ref{avg log transmission probability recurrence}) read
\begin{equation}
\avg{\ln |t_{1,n}|^2} \leq \avg{\ln |t_{1,n-1}|^2} + \ln |t_n|^2 - \frac{1}{N} \ln (1-|r_n|^N )^2
\end{equation}
and
\begin{equation}
\avg{\ln |t_{1,n}|^2} \geq \avg{\ln |t_{1,n-1}|^2} + \ln |t_n|^2 - \frac{1}{N} \ln (1+|r_n|^N )^2
\end{equation}
respectively. When applied repeatedly, these recurrences yield ($t_{1,1} \equiv t_1$)
\begin{align} \label{log avg transmission probability estimates}
\avg{\ln |t_{1,n}|^2} &\leq& \sum_{j=1}^n \ln |t_j|^2 - \frac{1}{N} \sum_{j=2}^{n} \ln (1-|r_j|^N)^2 ~,\nonumber\\
\avg{\ln |t_{1,n}|^2} &\geq& \sum_{j=1}^n \ln |t_j|^2 - \frac{1}{N} \sum_{j=2}^{n} \ln (1+|r_j|^N)^2 ~,
\end{align}
where we have omitted the lower index of the averaging brackets $\avg{\ldots}$.

We shall now assume that $t_j=t, r_j=r$ for all $j$. On the level of the Abelian anyonic quantum walk, this corresponds to a spatially independent coin. Exponentiating (\ref{log avg transmission probability estimates}) results in
\begin{align} \label{exp log avg transmission probability estimates}
\exp \avg{\ln |t_{1,n}|^2} &\leq& (1-|r|^N)^\frac{2}{N} e^{-n \left[\ln (1-|r|^N)^\frac{2}{N} - \ln |t|^2 \right]} ~,\nonumber\\
\exp \avg{\ln |t_{1,n}|^2} &\geq& (1+|r|^N)^\frac{2}{N} e^{-n \left[\ln (1+|r|^N)^\frac{2}{N} - \ln |t|^2 \right]} ~.
\end{align}
Estimates of the localization length $\xi_{loc}$ follow:
\begin{equation} \label{localization length estimatesII}
\frac{1}{\ln (1+|r|^N)^\frac{2}{N} - \ln |t|^2} \leq \xi_{loc} \leq \frac{1}{\ln (1-|r|^N)^\frac{2}{N} -\ln |t|^2} ~.
\end{equation}
For $N \rightarrow \infty$ we have $\xi_{loc} \rightarrow -\frac{1}{\ln |t|^2}$.

For the upper bound in (\ref{localization length estimates}) to make sense, $-\ln |t|^2 + \frac{1}{N} \ln (1-|r|^N)^2$ has to be a positive number. This leads to the condition
\begin{equation} \label{transmission reflection coefficient condition}
|t|^2 < (1-|r|^N)^{\frac{2}{N}} ~,~\textrm{i.e. }~ |t|^N + |r|^N < 1 ~.
\end{equation}
Since $|t|^2 + |r|^2 = 1$ (with $|t|,|r| < 1$), the latter is satisfied for $N>2$.

The case $N=1$ corresponds to fermions which are known not to localize (their exchange statistics does not induce any interference effects). The marginal case $N=2$ corresponds to \emph{semions} ($\varphi=\pi/2$), but we are unable to decide about their localization within this method.  However, numerics for an analogous model involving continuous time hopping of semions on a ladder support localisation \cite{ZMasters}.

To establish a connection between this scattering model and the Abelian anyonic quantum walk with the coin $U = \frac{1}{\sqrt{2}}\left( \begin{smallmatrix} 1&1\\1&-1 \end{smallmatrix} \right)$, we define
\begin{equation}
t = -\frac{1}{\sqrt{2}} ~,~ t' = \frac{1}{\sqrt{2}} ~,~ r = \frac{1}{\sqrt{2}} ~,~ r'= \frac{1}{\sqrt{2}} ~.
\end{equation}
The localization length estimate for $N=8$ ($\frac{\pi}{8}$-anyons) is now
\begin{equation}
1.412 \leq \xi_{loc} \leq 1.477 ~.
\end{equation}
Let us stress that we investigated stationary state of a wave after infinitely many scattering events. This corresponds to the infinite-time asymptotic behavior of the anyonic quantum walk.


\section{Simplification of the Jones polynomial $V_{L_{\vec{a}\vec{a}'}}(i)$}
The Jones polynomial for the links relevant to the quantum walk, can be related to a simpler $arf$ invariant through \cite{Murakami}
\begin{equation} \label{jones at point i}
V_{L_{\vec{a}\vec{a}'}}(i) = \sqrt{2}^{|\vec{m}|} (-1)^{arf(L_{\vec{a}\vec{a}'})} \prod_{\substack{s=1 \\ m_s>0}}^n \widetilde{\ell}_s ~,
\end{equation}
where $\widetilde{\ell}_s \equiv \delta_{0,\ell_s ~{\rm mod}~ 2}$. The product in the last expression is equal to $1$ only if the link $L_{\vec{a}\vec{a}'}$ is \emph{proper}, i.e. the sum of the pairwise linking numbers is even . Furthermore, when the link is \emph{totally proper}, i.e. all pairs of components have an even linking number, and there is no self-linking, then \cite{KirbyMelvin}
\begin{equation} \label{arf in local terms}
arf(L_{\vec{a}\vec{a}'}) = \sum_{s=1}^n m_s c_2(s) + \sum_{1 \leq s'<s'' \leq n} m_{s'} m_{s''} \tau(s',s'') ~,
\end{equation}
where $c_2(s)$ is the cubic coefficient in the \emph{Conway polynomial} of the two-component sublink of $L_{\vec{a}\vec{a}'}$ consisting of strands corresponding to the walker and island $s$; and $\tau(s',s'')$ is the \emph{Milnor triple point invariant} of the three-component sublink of $L_{\vec{a}\vec{a}'}$ consisting of the walker's strand and the strands corresponding to islands $s'$ and $s''$.  Note there is no self-linking because the links represent world lines of the anyons which move forward in time only.  In Ref. \cite{LehmanII} it was shown that for a uniformly filled background, for all the links that contribute to the diagonal elements of the spatial probability distribution, properness implies total properness.  For non-uniform filling the same is true for the simple reason that paths that had even linking between the walker and one island anyon will then have multiple pairwise linking when the island has multiple occupancy. 

Inserting (\ref{jones at point i}) and (\ref{arf in local terms}) into the expression for the fusion space trace (\ref{JonesDef}), we obtain
\begin{align}
\tr \mathcal{Y}_{\vec{a}\vec{a}'}  
&= (-i^{-\frac{3}{4}})^{w(L_{\vec{a}\vec{a}'})} (-1)^{arf(L_{\vec{a}\vec{a}'})} \prod_{\substack{ s=1 \\ m_s>0 }}^n \widetilde{\ell}_s \nonumber\\
&= (-1)^{\sum_{s'<s''} m_{s'} m_{s''} \tau(s',s'')}
\prod_{\substack{ s=1 \\ m_s>0 }}^n \widetilde{\ell}_s (i)^{\frac{\ell_s}{2} m_s} (-1)^{m_s c_2(s)} ~.
\end{align}
For islands $s$ such that $m_s > 0$ and $\widetilde{\ell}_s = 1$, i.e. $\frac{\ell_s}{2} \in \mathbb{Z}$, we can use a result of \cite{LehmanII}, and simplify
\begin{equation}
c_2(s) = \frac{\ell_s}{6} (\ell_s^2-1) = \frac{\ell_s}{2} \frac{1}{3} \left[ 4 \left(\frac{\ell_s}{2}\right)^2 - 1 \right] \overset{mod~2}{=} \frac{\ell_s}{2} ~.
\end{equation}

Hence
\begin{equation}
\tr \mathcal{Y}_{\vec{a}\vec{a}'} = \prod_{\substack{s=1 \\ m_s>0}}^n \widetilde{\ell}_s (-i)^{\frac{\ell_s}{2} m_s} \prod_{1 \leq s' < s'' \leq n} (-1)^{m_{s'} m_{s''} \tau(s',s'')} ~. \end{equation}

\section{Configuration average of $\tr \mathcal{Y}_{\vec{a}\vec{a}'}$ in Ising model anyonic quantum walk}

We assume uniform island occupation probabilities: $W(m) = 1/4$, $m=1,\ldots,4$. The average trace over the fusion space for Ising model anyons is calculated as
\begin{align}
\avg{\tr\mathcal{Y}_{\vec{a}\vec{a}'}} 
&= \sum_{\vec{m} \in \{1,\ldots,4\}^{n}} \frac{1}{4^n} \prod_{s=1}^{n} \widetilde{\ell}_s (-i)^{\frac{\ell_s}{2} m_s} \nonumber\\
&~~~~~~\times \prod_{1\leq s'<s'' \leq n} (-1)^{m_{s'} m_{s''} \tau(s',s'')} ~.
\end{align}
We observe the following:
\begin{align}
\avg{\tr\mathcal{Y}_{\vec{a}\vec{a}'}}
&= \left[ \prod_{s=1}^{n} \widetilde{\ell}_s \right] \frac{1}{4^{n-1}} \sum_{\vec{m} \in \{1,\ldots,4\}^{n-1}} \prod_{s=1}^{n-1} (-i)^{\frac{\ell_s}{2} m_s} \nonumber\\
&~~~~~~ \prod_{1\leq s'<s'' \leq n-1} (-1)^{m_{s'} m_{s''} \tau(s',s'')} \nonumber\\ 
&\times \frac{1}{4} \sum_{m_n=1}^4 \left[ (-i)^{\frac{\ell_n}{2}}\right]^{m_n} \left[ \prod_{s'=1}^{n-1} (-1)^{m_{s'} \tau(s',n)} \right]^{m_n} \nonumber\\
&= \ldots \times 
\left\{\begin{array}{ll}
1 & ~\textrm{if} ~ (-i)^{\frac{\ell_n}{2}}(-1)^{ \sum_{s'=1}^{n-1} m_{s'} \tau(s',n)} = 1 \\
0 & ~\textrm{otherwise}
\end{array}\right. ~
\end{align}
Hence, $\avg{\tr\mathcal{Y}_{\vec{a}\vec{a}'}} = 0$ whenever $\ell_n \neq 0 \textrm{ mod } 4$ (i.e. when $(-i)^{\frac{\ell_n}{2}}$ is not a real number). Furthermore, if $\ell_n = 4 \textrm{ mod } 8$, then $\tau(s',n)=0$ for all $r=1,\ldots,n-1$\footnote{Recall that $\tau(s',s'')$ can be nonzero only if $\ell_{s'},\ell_{s''}=0$.}, and therefore $(-1)^{\frac{\ell_n}{4}}(-1)^{ \sum_{s'=1}^{n-1} m_{s'} \tau(s',n)} = -1$. Hence, $\avg{\tr\mathcal{Y}_{\vec{a}\vec{a}'}} = 0$ whenever $\ell_n \neq 0 \textrm{ mod } 8$. By the same reasoning, analogous holds for any island $s=1,\ldots,n$.

Let us define $\widetilde{\widetilde{\ell_s}} \equiv \delta_{0,\ell_s ~{\rm mod}~ 8}$. We have
\begin{equation}
\avg{\tr\mathcal{Y}_{\vec{a}\vec{a}'}} = \displaystyle{ \prod_{s=1}^{n}} \widetilde{\widetilde{\ell_s}} \displaystyle{\sum_{\vec{m} \in \{1,\ldots 4\}^{n}} 4^{-n}(-1)^{\sum_{1\leq s'<s'' \leq n}m_{s'} m_{s''} \tau(s',s'')}}
\end{equation}
The expression $(-1)^{m_{s'} m_{s''} \tau(s',s'')}$ is invariant under shifting $m_{s'} \rightarrow m_{s'}+2$ or $m_{s''} \rightarrow m_{s''}+2$. Therefore, the sum over $\vec{m} \in \{1,\ldots,4\}^{n}$ contains $2^n$ classes of $2^n$ equivalent configurations. Also, $m_j=0$ is equivalent with $m_j=2$. We conclude
\begin{equation}
\avg{\tr\mathcal{Y}_{\vec{a}\vec{a}'}} = \left[ \prod_{s=1}^{n} \widetilde{\widetilde{\ell_s}} \right] \frac{1}{2^n} \sum_{\vec{m} \in \{0,1\}^{n}} \prod_{1\leq s'<s'' \leq n} (-1)^{m_{s'} m_{s''} \tau(s',s'')} ~.
\end{equation}

\section{Correlations in time}

To analyse the long-term behaviour of the non-Abelian case, we draw an analogy to
decoherent quantum walks by interpreting the anyonic quantum walk as a quantum walk with
an environment.
The fusion space is a very special kind of
environment, since it is non-local and highly non-Markovian, and such environments are not
well studied in the literature. Thus, anyonic environments open a new line of study on
effects of environment and decoherence in quantum systems.

Disordered quantum walks have been studied in the literature recently, and localization has been observed for
many types of spatial disorder in the coin parameter. Linden and Sharam \cite{spatialloc1} showed that
periodically varying coin parameters can lead to both bounded and unbounded walks, depending on the period.
Konno \cite{spatialloc2,spatialloc3} used path-counting methods to study quantum walks under quenched and
annealed disorder, and showed that a coin defect at the origin causes partial localization of the wavepacket.
Joye and Merkli \cite{spatialloc4} and Ahlbrecht et al \cite{spatialloc5} showed that dynamical Anderson
localization occurs for spatially inhomogeneous coin, when the coin parameters are chosen randomly from
continuous or certain discrete sets. If the coin parameters change in time, the walker spreads ballistically or
diffusively, but no localization has been observed.
Brun et al \cite{temporalloc1} studied periodically changing coins, and found that the walk is still ballistic
unless the period becomes as large as the length of the walk, in which case it is diffusive.
Shapira et al \cite{temporalloc2} studied quenched unitary noise numerically and observed that the walk
is diffusive in the long term after an initial ballistic period. Similar conclusions were drawn in analytical treatment
by Ahlbrecht et al \cite{temporalloc4} and  Joye \cite{temporalloc5}.

As shown in Ref. \cite{spatiotemporal},
the quantum walk is diffusive also in the presence of both temporal and spatial disorder,
in other words localization does not occur in a spatially disordered system if
temporal randomness is also present. The quantum coherent terms that are needed for localization
in the spatially disordered case are thus destroyed by decoherence from temporal randomness.
When a coin changes completely randomly, the identity of the next coin is not dependent on the previous coins at all,
and the environment
which induces the change has no memory. If the rate of change of the coin is larger than a single time step,
there exists a Markov process characterized by some time-varying parameter $\gamma(t)$. If there is
no memory at all, these constants are identical for every $t$. In a Markov process, the future states of a system
depend only on its present state and $\gamma$, not its history. In the following, we argue that the
anyonic environment is effectively memoryless and is therefore described by a Markov process,
which implies that the walker spreads diffusively.

Equation (\ref{average trace Ising}) in the main text shows that the average trace
$\langle\langle\tr\mathcal{Y}_{\vec{a}\vec{a}'}^{\rm (Ising)}\rangle\rangle$ of Ising anyons is almost identical
to that of Abelian anyons, except for the term involving the Milnor triple invariant $\tau$.
The Milnor triple invariant gives the number of Borromean rings in a three-component link.
We argue that the $\mathcal{T}_{\vec{a}\vec{a}'}$-term induces temporal randomness in the walk, such
that the environment acts
effectively in a Markovian way. The effect of this term is that it multiplies the contribution
from each path by  the configuration average over $(-1)^{\sum_{1\leq s'<s'' \leq n} m_{s'} m_{s''} \tau(s',s'')}$.
While this term preserves memory of the whole history of the particle's trajectory, we argue that
at short time scales it fluctuates in a disordered manner. The value of $\tau$ changes when a new
Borromean ring is formed, which requires at least 4 time steps. Also, the formation of a Borromean
ring requires a very specific pattern in the particle's trajectory which is in no way periodic.
In addition, because of the condition on the last coin outcomes $a_t=a_t'$, there are new path
patterns up to $t-1$ time steps introduced on every time step which were not allowed for the previous
time step, so these paths are not correlated to previous evolution at all.

\begin{figure}
\begin{center}
\includegraphics[width=\columnwidth]{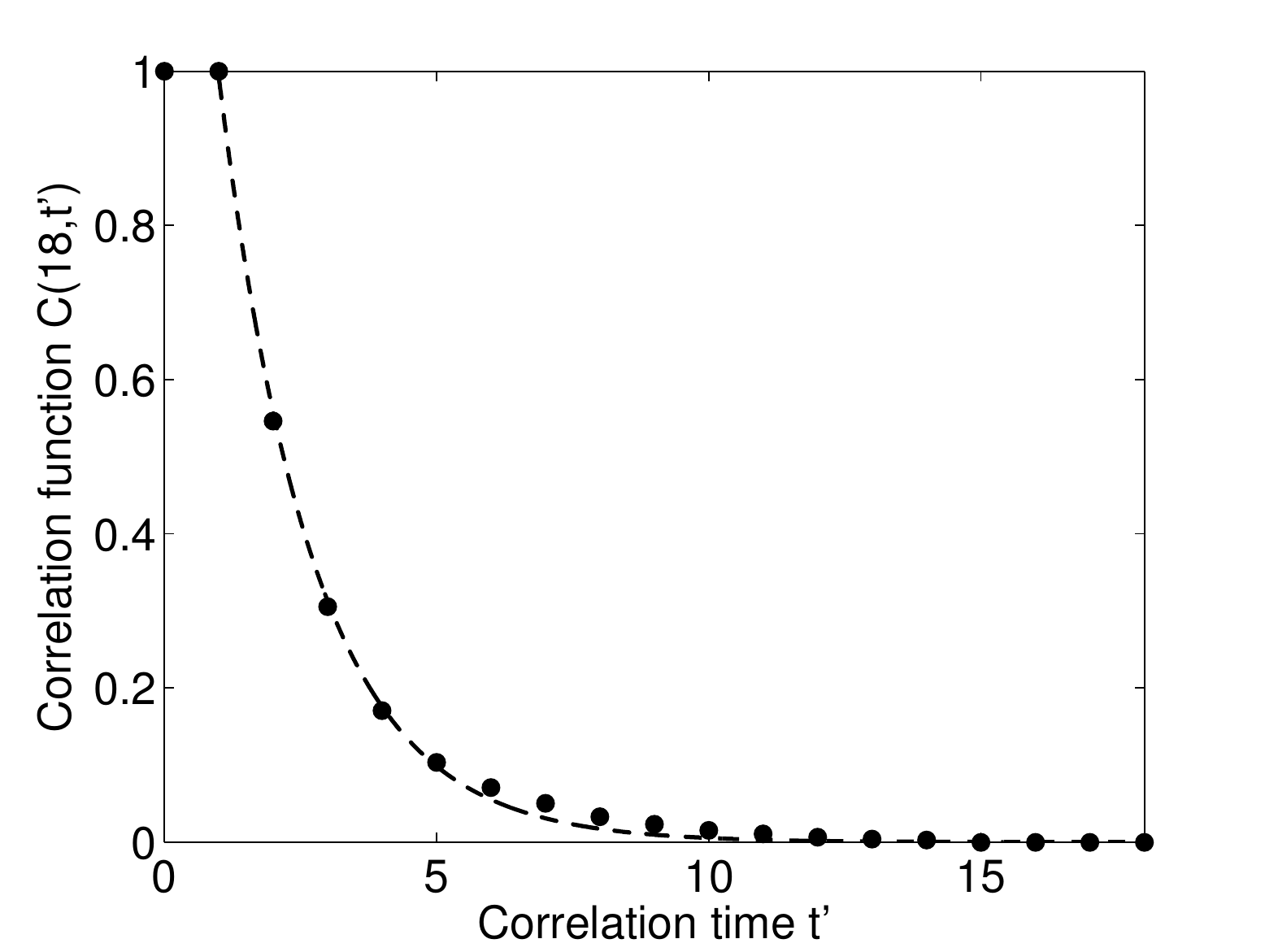}
\end{center}
\caption{Correlator $C(t,t')$ as a function of $t'$ with total number of time steps $t=18$.
The line shows the best exponential fit $C(t')=1.7679\:e^{-0.57699\,t'}$.}
\label{fig:taucorrelator}
\end{figure}

The time correlations of the $\tau$ invariant can be tracked by defining the correlator
\[
C(t,t') = \frac{\big\langle(-1)^{\tau_t}(-1)^{\tau_{t-t'}}\big\rangle
- \big\langle(-1)^{\tau_t}\big\rangle\big\langle
(-1)^{\tau_{t-t'}}\big\rangle}
{1-\big\langle(-1)^{\tau_t}\rangle^2}
\]
where $\tau_t=\sum_{1\leq s'<s'' \leq n} \tau(s',s'')$ is the sum of three-component invariants for all
sublinks for a path up to $t$ time steps and $\big\langle\cdot\big\rangle_{(\vec{a},\vec{a}')\leadsto s_0}$
is the expectation value over all paths leading to the initial site $s_0$ (the subindex has been suppressed
above for clarity). The term in the denominator
is a normalization factor, which is defined to be the value of the term in the numerator in the perfectly
correlated case $t'=0$. For simplicity, the
correlator is calculated only for the uniform filling ($m_s=1\quad\forall s$) using the same method as
described in \cite{LehmanII}. The time correlations can be analysed by keeping the final time
$t$ fixed and calculating the correlator for increasing values of $t'$. The intermediate time
value $\tau_{t-t'}$ is calculated by erasing the $t'$ last braid generators from the total braid
word, such that braiding is switched off after $t-t'$ time steps. For some braid words, the
quantity $\tau(s',s'')$ is not well defined in our method,
in which case we set $\tau(s',s'')=0$. The correlator for $t=18$ is plotted in Fig. \ref{fig:taucorrelator},
which shows that the correlations fall off exponentially after one step, indicating that effectively the $\tau$
invariant maintains memory only for a short period of time, and the environment is Markovian at
long time scales. Note that because of the condition $a_t=a_t'$, the braiding at the last time
step can always be trivially undone, therefore the last two time steps are perfectly correlated.
The correlator can also be calculated for other sites $s\neq s_0$. In these cases, the rapid
falloff is observed for the central region $s_0-t/\sqrt{2}\leq s\leq s_0+t/\sqrt{2}$, and outside this region
the falloff becomes linear at the furthest edge sites. The edge behaviour is however irrelevant,
as the behaviour of the quantum walk is determined by the central region only.

The fundamental reason for the classical-like behaviour of the walker is the strong entanglement
between the quantum walk states and the fusion states, and the highly mixing nature of the fusion
space environment. The walker states stay entangled with the fusion states for long time periods,
and recurrences where these states become uncoupled happen very rarely.

We have also investigated the effect of temporally random phases introduced to the spatially
random Abelian walk. In this model the wave function is multiplied by a random -1 phase with
some probability $p_{-1}$ if the walker crosses a site belonging to the temporally fluctuating
region of sites. The calculations with different values of $p_{-1}$ and different sizes of the
region up to 500 time steps showed that the behaviour becomes diffusive in all these cases.

\section{Numerical calculations of non-Abelian anyons}

We have calculated numerically the average variance $\avg{\sigma^2{^{\rm (Ising)}(t)}}$ over
configurations $m_s\in\{0,\ldots,4\}$
for the probability distribution given by Eqs. (\ref{fixed configuration distribution}) and (\ref{Ising fusion trace})
up to 23 time steps (46 anyons in the lattice), with at least 100 charge configurations.
The error bars were obtained by the standard deviation
of the variance of the spatial probability distribution. The average variance is approximately a straight
line with slope 0.456 from 10 to 23 time steps, but the error bars of the variance overlap with the errorbars of the
Abelian case, so we cannot distinguish between Abelian and non-Abelian anyons on this short time scale.
Considering only occupations $m_s\in\{1,\ldots,4\}$ (no vacuum charges), the variance is smaller, ie.
the wave packet is diffusing slower (not shown).


To obtain results for longer times, we turn to the continuous-time picture where the time evolution
is generated by a Hubbard-type Hamiltonian $H$ and the propagator $e^{-iH\delta t}$ implements an infinitesimal time evolution.
The total Hamiltonian is given by the sum of the shift and coin flip terms
$H=H_{\text{shift}}+H_{\text{flip}}$, where \cite{sukhi}
\[
H_{\text{shift}} = J\sum_s(T_{s+1}^-\hat{b}_sP_1+T_s^+\hat{b}_sP_2) + \text{h.c.}
\]
\[
H_{\text{flip}} = \sum_s(\kappa\ket{2}_s\bra{1}+\kappa^*\ket{1}_s\bra{2})
\]
with $J\in \mathbb{R}, \kappa\in \mathbb{C}$ and $T_s^\pm=(\ket{1}_{s\pm1}\;_s\bra{1}+\ket{2}_{s\pm1}\;_s\bra{2})\otimes
I_{\text{fusion}}$ are translation operators between sites $s$ and $s\pm1$,
$\hat{b}_s$ are braid generators as defined in the main text for braiding the mobile anyone around anyons in the island between $s$ and $s+1$ (acting on fusion space only) and
$P_c=\sum_s\ket{c}_s\bra{c}\otimes I_{\text{fusion}}$ are projectors to the coin states.
Here $\ket{c}_s$ corresponds to occupation of state $c$ at site $s$, i.e. $\ket{c=0}_s$ corresponds to no mobile anyon at site $s$, $\ket{c=1}_s(\ket{c=2}_s)$ is a mobile anyon with coin state $\ket{0}(\ket{1})$ at site $s$.
Here we consider only the case where coins are identical on every site.

The above Hamiltonian is the generator of continuous-time evolution for total time $T$. Running the continuous-time
walk for time $T$ simulates the discrete-time quantum walk in a stroboscopic manner, such that the walker
makes $T/\delta t$ steps of infinitesimal length $\delta t$: $e^{-iHT}=(e^{-iH\delta t})^{T/\delta t}$. In the first order of Suzuki-Trotter expansion, the propagator decomposes
to $e^{-iH_{\text{shift}}\delta t}e^{-iH_{\text{flip}}\delta t}$, similarly as  the single step operator
in the discrete-time quantum walk model.

The TEBD algorithm was used to perform real time evolution of an initial state with the mobile anyon placed at the middle of the lattice and with couplings $\kappa=J=1$. Our implementation of the TEBD algorithm explicitly preserves anyonic charge \cite{anyonicTensorNetworks} and also particle number \cite{Singh11} corresponding to the presence of a single walker. Thus, while the position of the walker and the fusion degrees of freedom of the anyons are entangled during the time evolution, the algorithm preserves distinction between them. This in turn allowed for specification of total anyonic charge and particle number, here we considered the total vacuum sector with one walker.


\begin{thebibliography}{25}
\bibitem{footnote}
If we include the vacuum island occupation and consider a model with $W(m)=1/5$ for $1 \leq m \leq 5$, numerics show that the variance of the resulting walker's position distribution grows even faster.
\end{thebibliography}
\end{document}